\documentclass[10pt,leqno]{amsart}
\usepackage{graphicx}
\usepackage{indentfirst,csquotes}

\usepackage{amssymb,amsthm,amsmath}
\usepackage{xcolor,paralist,hyperref,titlesec,fancyhdr,etoolbox}

\titleformat{\section}[display]{\normalfont\huge\bfseries\centering}{\centering\chaptertitlename\thechapter}{10pt}{\Large}
\titlespacing*{\section}{0pt}{0ex}{0ex}

\hypersetup{ colorlinks=true, linkcolor=black, filecolor=black, urlcolor=black }

\usepackage{lipsum}

\usepackage{amsfonts}
\usepackage{subcaption}
\usepackage{algorithmic}
\usepackage{graphicx,wrapfig}
\usepackage{tabularx}
\usepackage{siunitx}
\usepackage{textcomp}
\usepackage{array} 
\usepackage{float}
\usepackage{adjustbox}
\usepackage{url}\usepackage{lscape}
 \usepackage{xspace}
\usepackage{makecell}
\usepackage{listings}
\usepackage{balance}
\usepackage{enumitem}
\usepackage{multirow}
\usepackage{rotating}
\usepackage{booktabs}
\usepackage{tikz}
\usepackage{pgfplots, pgfplotstable}
\pgfplotsset{
   compat=1.11,
   /pgf/number format/.cd,use comma,
   1000 sep = {\,},
       min exponent for 1000 sep = 4}
\usepackage{txfonts} 
\usetikzlibrary{positioning, shapes.geometric, arrows.meta}
\usepgfplotslibrary{groupplots}
\usetikzlibrary{patterns}

\lstdefinestyle{mystyle}{
    backgroundcolor=\color{white},   
    commentstyle=\color{green},      
    keywordstyle=\color{blue},       
    numberstyle=\tiny\color{gray},   
    stringstyle=\color{red},         
    basicstyle=\ttfamily\scriptsize,       
    breakatwhitespace=false,         
    breaklines=true,                 
    captionpos=b,                    
    keepspaces=true,                 
    numbers=left,                    
    numbersep=5pt,                   
    showspaces=false,                
    showstringspaces=false,          
    showtabs=false                   
}

\pgfplotsset{
   compat=1.11,
   /pgf/number format/.cd,use comma,
   1000 sep = {\,},
       min exponent for 1000 sep = 4}
\usepackage{txfonts}

\definecolor{mylightgray}{RGB}{224,224,224}

\newboolean{showcomments}
\setboolean{showcomments}{true}
\ifthenelse{\boolean{showcomments}}

\setboolean{showcomments}{true}
\ifthenelse{\boolean{showcomments}}
 { }

\newcommand{\finding}[1]{
	\setlength{\fboxrule}{1pt}
	\begin{center}\
		\noindent\fcolorbox{black}{gray!10}{
		\begin{minipage}{.92\linewidth}
			#1
		\end{minipage}
	}
	\end{center}
	\smallskip
}

\usepackage{pifont}
\newcommand{\wcircle}[1]{\ding{\numexpr171 + #1}}

\input{yaml}

\makeatletter
\renewcommand\section{\@startsection{section}{1}{\z@}%
  {-3.5ex plus -1ex minus -.2ex}
  {2.3ex plus .2ex}
  {\normalfont\Large\bfseries\centering}} 

\renewcommand\subsection{\@startsection{subsection}{2}{\z@}%
  {-2.5ex plus -1ex minus -.2ex}%
  {1.5ex plus .2ex}%
  {\normalfont\large\bfseries}} 
\makeatother

\begin{document}
\title{Security Smells in Infrastructure as Code Scripts: a Taxonomy Update Beyond the Seven Sins} 

\author[]{
Aicha WAR
\and
Serge L. B. NIKIEMA
\and
Jordan Samhi
\and
Jacques KLEIN
\and
Tegawendé F. BISSYANDE \\
University of Luxembourg \\
}

\maketitle

\begin{abstract}
Infrastructure as Code (IaC) has become essential for modern software management, yet security flaws in IaC scripts can have severe consequences, as exemplified by the recurring exploits of Cloud Web Services. Prior work has recognized the need to build a precise taxonomy of security smells in IaC scripts as a first step towards developing approaches to improve IaC security. 
This first effort led to the unveiling of seven sins, limited by the focus on a single IaC tool as well as by the extensive, and potentially biased, manual effort that was required. 
We propose, in our work to, revisit this taxonomy: first, we extend the study of IaC security smells to a more diverse dataset with scripts associated with seven popular IaC tools, including Terraform, Ansible, Chef, Puppet, Pulumi, Saltstack, and Vagrant;
second, we bring in some automation for the analysis by relying on an LLM. While we leverage LLMs for initial pattern processing, all taxonomic decisions underwent systematic human validation and reconciliation with established security standards.
Our study yields a comprehensive taxonomy of 62 security smell categories, significantly expanding beyond the previously known seven.

We demonstrate actionability by implementing new security checking rules within linters for seven popular IaC tools, often achieving 1.00 precision score.
Our evolution study of security smells in GitHub projects reveals that these issues persist for extended periods, likely due to inadequate detection and mitigation tools. This work provides IaC practitioners with insights for addressing common security smells and systematically adopting DevSecOps practices to build safer infrastructure code.
\end{abstract}



\section{Introduction} 
\label{sec:introduction}

The widespread adoption of DevOps~\cite{ebert2016devops} has established Infrastructure as Code (IaC) as a key approach to software management. IaC automates software management tasks including installations, configurations, provisioning, and deployments across servers, operating systems, virtual machines, containers, and applications in both cloud and on-premises environments~\cite{morris2020infrastructure}.

As IaC practices become widespread, ensuring script security is essential to protect underlying IT infrastructure. IaC scripts are vulnerable to security issues such as misconfigurations~\cite{lepiller2021analyzing}, inadvertent secret exposures~\cite{9652648}, lax access controls, and resource leaks~\cite{8812041, 9388795}. Detecting, understanding, and mitigating these issues is paramount for safeguarding deployed systems.

Script coding issues often manifest as recurring patterns known as \textbf{code smells}~\cite{8590193, 9388795}, associated with maintainability issues and potential infrastructure management problems. \textbf{Security smells}, a specific type of code smell, are recurrent patterns signaling security vulnerabilities within code~\cite{10102545, 10174011, 10.1145/3377811.3380409, 10.1145/3408897, 9388795}. 

Rahman \textit{et al}.~\cite{8812041} recognized the need for a precise taxonomy of security smells in IaC scripts as a foundation for developing security improvement approaches, unveiling seven categories. However, this seminal work has two main limitations: first, it focuses on a single IaC tool (Puppet), excluding popular tools from the practitioner community such as Chef, Terraform, and Ansible (for instance, Terraform is used by 10 times more professionals than Puppet\footnote{\url{https://survey.stackoverflow.co/2024/technology\#most-popular-technologies}}). Second, the taxonomy relied entirely on manual effort, potentially introducing biases and comprehensiveness limitations.

{\bf This paper} revisits this taxonomy by extending IaC security smell studies to seven popular tools performing diverse automation tasks: Ansible, Terraform, Chef, Puppet, Pulumi, Saltstack, and Vagrant.

We employ Large Language Models (LLMs)~\cite{ye2023comprehensive, fang2024large} as assistive tools—under systematic human supervision—to help process the large-scale analysis of security patterns across seven IaC tools. While LLMs aided in initial clustering and pattern recognition, every categorization underwent rigorous human validation, reconciliation against established security standards (CWE), and external review. This human-in-the-loop approach, combined with the broader scope of seven IaC tools (versus one in prior work), enabled us to expand the taxonomy substantially.
Through this new experimental setup, our taxonomy has expanded substantially over existing work.
Through systematic analysis of seven diverse IaC tools—compared to the single tool in prior work—and careful human validation of LLM-assisted categorization, we identify 62 fine-grained security smell categories, substantially expanding the original 7 coarse-grained categories.

Based on our taxonomy, we augment detection rule sets for seven IaC tool linters to help practitioners address security smells. Unlike prior work implementing security rules in custom linters for specific tools, we enhance existing linters across seven popular IaC tools with consistent security rule sets.

The main contributions of our study are:
\begin{itemize}[noitemsep,topsep=0pt]
\item Human-validated identification of 62 categories of security smells impacting IaC scripts, derived through LLM-assisted analysis with systematic quality controls, expanding previous work by examining seven popular IaC tools (Ansible, Terraform, Chef, Puppet, Pulumi, Vagrant, and Saltstack).
\item Empirical analysis of security smell persistence in IaC across projects. 
\item Enhanced linters for seven IaC tools with static analysis rules detecting the 10 most representative security smells. 
\end{itemize}

We emphasize that while LLMs accelerated our analysis, they served strictly as tools under human control. The taxonomy's validity rests on multi-layered validation: cross-referencing with CWE standards, manual reconciliation by multiple researchers, and external annotator verification.

\section{Background and Related Work}
\label{sec:background}
In this section we explore our study background as well as previous studies that are similar or linked to our research. We highlight our contributions on top of this related work.
\subsection{Background}

\noindent
\textbf{DevOps and DevSecOps}.
DevOps refers to the automation of development (Dev) tasks such as unit tests, integration tests, etc., as well as the automation of operation (Ops) tasks such as artifacts releases or deployments~\cite{colakoglu2021software}. 
This automation fits into  three main pipelines:
Continuous Integration (CI) for development tasks, Continuous Delivery (CDE) for releases, and Continuous Deployment (CD) for IT operation tasks. Our study focuses on CD, specifically on the security aspects of Infrastructure as Code. To achieve a secure DevOps workflow, security practices and collaboration with security operators have been taken into account within processes referred to as DevSecOps or SecDevOps~\cite{9781709, myrbakken2017devsecops}. Despite possible nuances, in this paper, we refer to DevSecOps as an umbrella term for both DevSecOps and SecDevOps. We also use the terms ``IaC maintainers'' and ``IaC practitioners'' as umbrella terms for developers, IT operators, and DevOps engineers.


\noindent
\textbf{Infrastructure as Code.}
IaC is a practice that relies on code, including scripts and predefined components, to automate the provisioning and management of IT infrastructure, streamlining deployment and reducing human error~\cite{morris2020infrastructure}. 

 Figure \ref{fig:IaC} provides a code snippet from a script written using the Ansible IaC tool for the attribution of rights to a guest user. This is an example of a vulnerable script that assigns overprivileged roles to a guest user and performs the following actions :
 \begin{itemize}[noitemsep,topsep=0pt,left=3pt]
    \item The ``ansible.builtin.user'' module (lines 7--11) ensures that guest user exists on the target system.
    \item The ``ansible.builtin.copy'' module (lines 14--17) creates a sudoers file for the guest user, granting it passwordless sudo privileges.
    \item The ``ansible.builtin.group'' module (lines 20--22) ensures that a group for the guest role exists.
    \item The ``ansible.builtin.user''  module (lines 25--28) is then used again to add the guest user to the guest role group.
 \end{itemize}

 \begin{figure}[h]
    \centering
\begin{adjustbox}{max width=.7\linewidth}
\scriptsize
  \begin{lstlisting}[language=yaml, numbers=left]
---
- name: Assign role to guest user
  hosts: all
  become: yes
  tasks:
    - name: Ensure guest user exists
      ansible.builtin.user:
        name: guest
        comment: "Guest User"
        shell: /bin/bash
        state: present

    - name: Assign role to guest user
      ansible.builtin.copy:
        dest: /etc/sudoers.d/guest #Role over assignment to guest user
        content: |
          guest ALL=(ALL) NOPASSWD:ALL #Grant passwordless sudo privileges

    - name: Create a group for guest role
      ansible.builtin.group:
        name: guest_role
        state: present

    - name: Add guest user to guest_role group
      ansible.builtin.user:
        name: guest
        groups: guest_role
        append: yes

\end{lstlisting}
\end{adjustbox}
     \caption{Ansible Script Example: Over Assigning a Role to a Guest User}
     \label{fig:IaC}
 \end{figure}

IaC scripts can seamlessly manipulate resources like files, utilize modules from plugins, manage user accounts, and perform various system configurations, enabling efficient and automated infrastructure management as described in the previous example. Code security smells usually happen when IaC practitioners manipulate these script components.

Prior research on security smells in IaC primarily focused on specific configuration management tools like Ansible, Chef and Puppet~\cite{van2018good,sharma2016does,rahman2018characterizing}. This limited the scope of potential security smells, as IaC encompasses a broader range of tools. Indeed, IaC tools can be categorized into two groups depending on their main tasks. \ding{182} Orchestration tools (Terraform and Vagrant, for example) that perform provisioning, organizing, and managing processes for infrastructure components. \ding{183} Configuration management tools (Ansible, Chef, Puppet, for example) that allow installations, updates, or management of softwares in infrastructure components ~\cite{morris2020infrastructure}. In our work, we consider both types of tools as new directions for investigating security smells in IaC scripts compared to prior studies. Our research aims to provide a more comprehensive assessment of security vulnerabilities within the IaC ecosystem. 





\subsection{Related Work}

\noindent
\textbf{Taxonomy of security smells in IaC.}
Prior to 2018, research on security practices within Infrastructure as Code was scarce. Existing studies primarily focused on code maintainability for Chef and Puppet scripts~\cite{van2018good,sharma2016does,rahman2018characterizing} and identifying code smells within IaC~\cite{8590193}.

Since then, significant progress has been made in the field of IaC security. Researchers have developed methods to categorize~\cite{8812041, 9388795}, characterize ~\cite{10.1145/3417113.3422154}, and detect ~\cite{10.1145/3408897} security smells in IaC scripts.

These efforts were followed by other studies that established a taxonomy of security vulnerabilities within IaC scripts~\cite{10.1145/3377811.3380409}. Furthermore, recent advances highlight the potential of AI models, often coupled with static analysis tools, for predicting defects in IaC~\cite{10516612, 8367077, 10.1145/3405962.3405979, 10.1145/3416505.3423564}. Our work aligns with this growing trend, focusing on detecting security smells in IaC scripts. Our approach fills a gap and overlooks the wider landscape of IaC scripts encompassing a diverse set of tools. We mainly differentiate ourselves by conducting, beyond Ansible, Chef and Puppet scripts, a comprehensive study on seven types of scripts from different families of IaC automation tools as potential sources of security smells. Additionally, because previous studies on security smells in IaC scripts were heuristic-based, we innovate and extend current knowledge by incorporating an LLM to automatically identify and categorize security smells in our collected scripts.

\noindent
\textbf{Large Language Models.}
In recent years, Large Language Models (LLMs) such as OpenAI's GPT have gained significant traction in the realm of software development, offering innovative solutions for enhancing coding practices and improving software quality~\cite{wadhwa2023frustrated, spiess2024quality}. These models utilize advanced natural language processing techniques to understand~\cite{10.1145/3597503.3639187} and generate code~\cite{e25060888}, enabling developers to receive context-aware~\cite{baek2024knowledge} code suggestions, automate documentation, and generate code snippets from natural language descriptions. Furthermore, LLMs have demonstrated efficiency in detecting code defects and security smells by analyzing code patterns and structures~\cite{yang2024large, jin2023inferfix}. Their ability to learn from extensive datasets allows them to identify recurring patterns that may indicate vulnerabilities or maintenance issues, thereby facilitating the detection of potential security weaknesses. All these successful efforts in the literature to test LLM capabilities and performances in natural language and code processing demonstrate that LLMs can be useful tools to help automate usual manual-based and heuristic-based investigations. Therefore, we leverage GPT-3.5-Turbo and GPT-4o, widely investigated in the literature and proficient in code review capabilities ~\cite{pornprasit2024gpt,bae2024enhancing} in our work to automatically identify and categorize security smells in our dataset of IaC scripts.

\section{Methodology}
\label{sec:methodology_review}


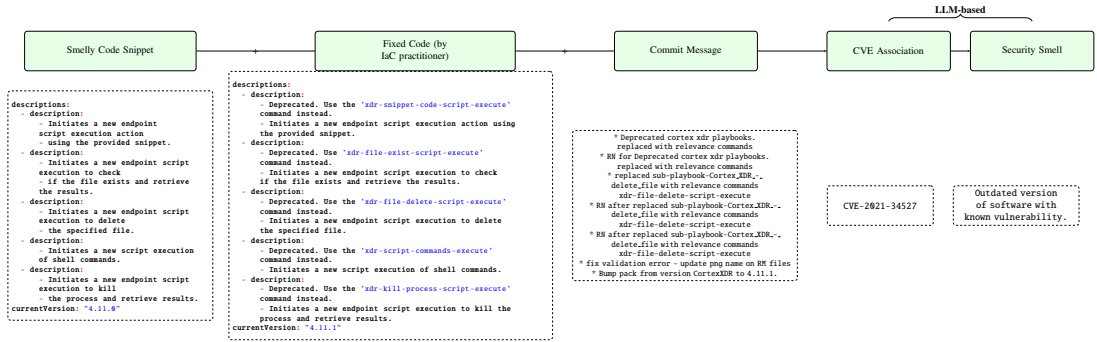
\begin{figure*}[t!]
\centering
\begin{adjustbox}{max width=\linewidth}
\begin{tikzpicture}[
    every node/.style={font=\LARGE}, 
    box/.style={draw, rounded corners, align=center, text width=6cm, minimum height=2cm, fill=green!10}, 
    dashedbox/.style={font=\large, draw, dashed, rounded corners, text width=3cm, align=center, anchor=north west, inner sep=5pt, minimum height=2cm},
    arrow/.style={thick, ->, >=stealth}
]

\newcommand{\boxwidth}{0.3\linewidth}  

\node (codesnippet) [box, minimum width=.6\linewidth] {Smelly Code Snippet};
\node (rawtext) [box, minimum width=.7\linewidth, right=6cm of codesnippet] {Fixed Code (by IaC practitioner)};
\node (rawtext2) [box, minimum width=.5\linewidth, right=5cm of rawtext] {Commit Message};
\node (initialcategory) [box, minimum width=.15\linewidth, right=3.5cm of rawtext2] {CVE Association};
\node (securitysmell) [box, minimum width=.15\linewidth, right=1cm of initialcategory] {Security Smell};

\draw (codesnippet)  -- node[midway] {\LARGE +}  (rawtext);
\draw (rawtext)  -- node[midway] {\LARGE +}  (rawtext2);

\draw [decorate, decoration={brace, amplitude=10pt}, thick]
    ([yshift=0.5cm]initialcategory.north) -- ([yshift=0.5cm]securitysmell.north)
    node[midway, yshift=0.6cm] {\textbf{LLM-based}};
\draw [arrow] (rawtext2) -- (initialcategory);
\draw [arrow] (initialcategory) -- (securitysmell);

\node (snippet1) [dashedbox, text width=10cm, below=1cm of codesnippet] {
    \begin{lstlisting}[language=yaml]
descriptions:
  - description: 
      - Initiates a new endpoint
      script execution action
      - using the provided snippet.
  - description: 
      - Initiates a new endpoint script 
      execution to check
      - if the file exists and retrieve 
      the results.
  - description: 
      - Initiates a new endpoint script 
      execution to delete
      - the specified file.
  - description: 
      - Initiates a new script execution 
      of shell commands.
  - description: 
      - Initiates a new endpoint script 
      execution to kill
      - the process and retrieve results.
currentVersion: "4.11.0"
\end{lstlisting}
};

\node (raw1) [dashedbox, text width=16cm,  below=5cm of rawtext, right=.8cm of snippet1] {
    \begin{lstlisting}[language=yaml]
descriptions:
  - description: 
      - Deprecated. Use the 'xdr-snippet-code-script-execute'
      command instead.
      - Initiates a new endpoint script execution action using
      the provided snippet.
  - description: 
      - Deprecated. Use 'xdr-file-exist-script-execute'
      command instead.
      - Initiates a new endpoint script execution to check 
      if the file exists and retrieve the results.
  - description: 
      - Deprecated. Use the 'xdr-file-delete-script-execute'
      command instead.
      - Initiates a new endpoint script execution to delete 
      the specified file.
  - description: 
      - Deprecated. Use the 'xdr-script-commands-execute'
      command instead.
      - Initiates a new script execution of shell commands.
  - description: 
      - Deprecated. Use the 'xdr-kill-process-script-execute'
      command instead.
      - Initiates a new endpoint script execution to kill the
      process and retrieve results.
currentVersion: "4.11.1"
\end{lstlisting}
};
\node (raw2) [dashedbox, style={font=\Large}, text width=11cm,  below=3cm of rawtext, right=1cm of raw1] {
    \texttt{* Deprecated cortex xdr playbooks. replaced with relevance commands}\\
    \texttt{* RN for Deprecated cortex xdr playbooks. replaced with relevance commands}\\
    \texttt{* replaced sub-playbook-Cortex\_XDR\_-\_\\delete\_file with relevance commands xdr-file-delete-script-execute}\\
    \texttt{* RN after replaced sub-playbook-Cortex\_XDR\_-\_\\delete\_file with relevance commands xdr-file-delete-script-execute}\\
    \texttt{* RN after replaced sub-playbook-Cortex\_XDR\_-\_\\delete\_file with relevance commands xdr-file-delete-script-execute} \\
    \texttt{* fix validation error - update png name on RM files}\\
    \texttt{* Bump pack from version CortexXDR to 4.11.1.}  \\
};

\node (category1) [dashedbox, style={font=\LARGE}, text width=5cm, below=3cm of initialcategory, right=1.5cm of raw2] {
    \texttt{CVE-2021-34527}\\
};

\node (smell1) [dashedbox, style={font=\LARGE}, text width=6cm, below=3cm of securitysmell, right=1cm of category1] {
    \texttt{Outdated version of software with known vulnerability.}
};

\end{tikzpicture}
\end{adjustbox}
\caption{An example of how we use an LLM to analyze and determine security smells based on CWE categories in an Ansible Playbook script.}
\label{fig:cwe}
\end{figure*}

\begin{figure*}[t!]
\centering
\begin{adjustbox}{max width=\linewidth,scale=1}
\begin{tikzpicture}[
    every node/.style={font=\large}, 
    box/.style={draw, rounded corners, align=center, text width=3cm, minimum height=1.5cm, fill=green!10}, 
    dashedbox/.style={draw, dashed, rounded corners, text width=3cm, align=center, anchor=north west, inner sep=5pt, minimum height=2cm},
    arrow/.style={thick, ->, >=stealth}
]

\newcommand{\boxwidth}{0.2\linewidth}  

\node (codesnippet) [box, minimum width=.1\linewidth] {Smelly Code Snippet};
\node (rawtext) [box, minimum width=.2\linewidth, right=0.3cm of codesnippet] {Fixed Code (by IaC practitioner)};
\node (description) [box, minimum width=.6\linewidth, right=4cm of rawtext] {LLM smell description};
\node (securitysmell) [box, minimum width=\boxwidth, right=3.5cm of description] {Security Smell};
\node (globalcategory) [box, minimum width=\boxwidth, right=1cm of securitysmell] {Global Security Category};

\draw (codesnippet) -- node[midway] {\LARGE +} (rawtext);
\draw [arrow] (rawtext) -- (description);
\draw [decorate, decoration={brace, amplitude=10pt}, thick]
    ([yshift=0.5cm]securitysmell.north) -- ([yshift=0.5cm]globalcategory.north)
    node[midway, yshift=0.6cm] {\textbf{LLM-based}};
\draw [arrow] (description) -- (securitysmell);
\draw [arrow] (securitysmell) -- (globalcategory);

\node (snippet1) [dashedbox,style={font=\small}, text width=2.4cm, below=1cm of codesnippet] {
    \begin{lstlisting}[language=yaml]
no_log: true
\end{lstlisting}
};

\node (raw1) [dashedbox,style={font=\small}, text width=3.6cm, below=1cm of rawtext, right=0.3cm of snippet1] {
 . 
    \begin{lstlisting}[language=yaml]
no_log: 
hide_sensitive_logs
\end{lstlisting}
};

\node (llm1) [dashedbox, style={font=\large}, text width=12cm, align=left, below=1cm of description.south ] {
    \texttt{The security smell associated with the previous code is that it uses the "no\_log: true" option, which completely disables logging for a specific task.}\\
    \texttt{This can be dangerous as it prevents sensitive information, such as passwords or API keys, from being logged during execution. }\\
    \texttt{If there is a need to troubleshoot or audit the playbook, having these logs can be crucial for identifying issues or unauthorized access.}\\
    \texttt{By using "no\_log: hide\_sensitive\_logs" instead, sensitive information will still be logged but will be masked or hidden to prevent it from being exposed in clear text.}\\
    \texttt{This allows for better security practices while still providing the necessary logging information for troubleshooting and auditing purposes.}
};

\node (smell1) [dashedbox, style={font=\large}, below=2cm of securitysmell, right=1cm of llm1] {
    \texttt{Insecure Logging}
};

\node (globalcategory1) [dashedbox, style={font=\large}, below=2cm of globalcategory, right=1cm of smell1] {
    \texttt{Logging and Monitoring}
};

\end{tikzpicture}
\end{adjustbox}
\caption{An example of how we use an LLM to determine security smells in an Ansible Playbook script.}
\label{fig:llm}
\end{figure*}
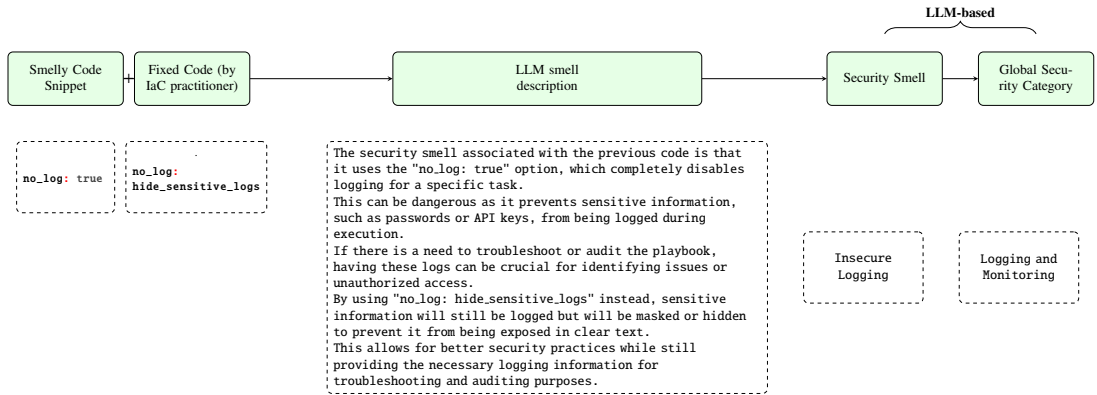
To comprehensively investigate security smells within IaC scripts, we first identified the most widely\footnote{Based on the StackOverflow survey and mentions in the scientific research literature.} used IaC tools, and then we collected a large set of scripts written by developers for these IaC tools. Within the project development history of those scripts, we rely on commit messages to identify security-fix commits based on keyword matching. Based on the security fix commit information, we collect the relevant code snippets for our study. To validate the collection of relevant code snippets, we further apply the Snyk tool~\cite{singh2022comparative,sushma2023detect}, which implements static rules and AI models to yield security warnings within IaC scripts. Given the final consolidated dataset, we then leverage an LLM to automate the clustering and labeling of security smell categories. In the remainder of this section, we provide details on the research questions that we investigated in our study, the data collection (cf. Section~\ref{sec:collection}), the methodology for automating the construction of the taxonomy (cf. Section~\ref{sec:taxonomy}), the analysis process of the evolution of security smells (cf. Section~\ref{sec:smells}) as well as the design of the static checking rules to augment existing linters for IaC scripts (cf. Section~\ref{sec:rules}).


Our study aims to answer three research questions that we describe in the following subsections.

\subsection{RQ1: What security smells exist within IaC scripts?}
This question identifies and categorizes prevalent security code smells within IaC scripts associated with a diverse set of IaC tools that target configuration, management and orchestration.

\subsubsection{Data Collection}
\label{sec:collection}
Our goal is to collect a dataset of code snippets affected by security smells in IaC scripts. 
To that end, we employ a systematic process relying on code changes that IaC practitioners committed to fix security issues.
Since closed-source projects used in practice are inaccessible, our study focuses exclusively on open-source projects available on GitHub.
To identify relevant commits, we used the GitHub API and searched for commits using security-related keywords\footnote{The full list of keywords is available in the replication package. Examples include `\textit{fix security}', `\textit{CVE}', etc.} including the names of the seven tools and their script types (e.g., cookbook, vagrantfile). This process yielded \num{2384} security-fix commits from the top results provided by the GitHub API for scripts associated with all seven tools. We then extracted script versions before the commits were applied to fix security issues using their commit SHA, and curated these scripts by focusing on those exhibiting security smells according to Snyk's popular static security analyzer for IaC. Snyk includes a large database of CVEs\footnote{Snyk CVE database--\url{https://security.snyk.io}} associated with our selected IaC tools.
This curation validates the final dataset and enables focus on precise code snippets where security warnings are highlighted.

After validation, we randomly and manually selected a final dataset of 1050 verified code snippets with security smells: 150 code snippets for each script type. This number is the minimum we could find for Chef and Vagrant.



\subsubsection{Categorization of Security Smells.}
\label{sec:taxonomy}
We describe the process for identifying, labeling, and categorizing security smells in IaC code snippets across various tools. A major novelty of our work is implementing automation using LLMs for analysis and category label generation. We leverage the fact that many LLMs are trained on open-source data (such as GitHub projects) to perform categorization of security smells.

We consider two distinct methods whose outputs are reconciled to obtain final categories, instilling greater confidence in the produced taxonomy. The first method is \textbf{extractive}, borrowing smell categories from existing labels (Common Weakness Enumerations - CWEs), while the second is \textbf{generative}, using AI-generated smell labels. Our extractive method serves as validation for the generative method, which constitutes our main contribution.

\vspace{0.2cm}
\noindent
{\ding{182}  \bf A CWE-oriented extractive categorisation}.
Figure~\ref{fig:cwe} illustrates our first categorization method, where the security smell categories are borrowed from existing terminology of CWE. The following steps outline the procedure for our first method, accompanied by an example to illustrate each step in detail.
\begin{itemize}[leftmargin=*]

\item \textit{Input preparation -}
The categorization is driven by the following triplet data as input: a smelly code snippet, the associated fixed version based on an IaC practitioner's commit, and the commit message that describes the implemented code change. 

\item \textit{CVE as output generated by the LLM -} Based on the input we apply the following prompt to the LLM: 

{\footnotesize
\begin{verbatim}
Prompt >>
    "Identify the relevant CVE for the following 
    [SCRIPT_TYPE] code snippet labeled as vulnerable.
    The context is as follows:
    Vulnerable code: [VULNERABLE_CODE]
    Fixed code: [FIXED_CODE]
    Commit message: [COMMIT_MESSAGE]"
>>>
\end{verbatim}
\noindent\textit{Note: The elements put inside "[]" are variables. The variable [SCRIPT\_TYPE] can have values like Ansible playbook, Chef cookbook, etc.}\\
}

As suggested by the prompt, we use the LLM to find CVEs associated with security issues addressed by IaC practitioners' code changes. LLMs trained with code corpora offer an effective engine to automate such lookups, as manual analysis would require tremendous effort when input data cannot be matched directly with CVE details.
 In the example of Figure \ref{fig:cwe}, the missing CVE details from the input did not prevent the LLM from suggesting the CVE number, which is then manually validated as being relevant in all our dataset.

\item \textit{Security Smell Categorization -} Once a CVE number is yielded, we can readily associate the CWE category. We refer to that category as the security smell, e.g., \textit{Outdated Software Version} in the illustrative example of Figure \ref{fig:cwe}.

\end{itemize}

\vspace{0.2cm}
\noindent
{\ding{183} \bf An LLM-based generative categorisation}. 
Our second method (cf. Figure~\ref{fig:cwe}) generates categories directly from LLM-based code analysis, following these steps:
\begin{itemize}[leftmargin=*]
\item \textit{Input preparation -}   
The context provided to the LLM consists of two code snippets: the smelly code and its fixed version. We exclude commit messages since they may contain CVE or CWE information directly used in the first method, avoiding bias toward common vulnerability descriptions.

\item \textit{Detailed smell description generation} 
We apply the input information to the following prompt:

{\footnotesize
\begin{verbatim}
Prompt >>>
    "Please provide a description of the security 
    smell associated with the following [SCRIPT_TYPE] 
    code snippet labeled as vulnerable. 
    We also provide its fixed version:
    Vulnerable code: [VULNERABLE_CODE], 
    Fixed code: [FIXED_CODE]"
>>>
\end{verbatim}
}

Following the prompt above, the objective is to prompt the LLM to generate a detailed explanation of the security smell associated with the provided smelly code, as in the example of Figure~\ref{fig:llm}.

\item \textit{Security smell category label generation -} We use the following prompt for the category labeling  of security smells:

{\footnotesize
\begin{verbatim}
prompt >>>
    "I want to build a taxonomy of security smells.
    Make clusters of similar security smell descriptions
    and provide categories of security smells for the
    following description(s): [SMELL_DESCRIPTION]" 
>>>
\end{verbatim}
}

As suggested by the description of the prompt above, the LLM is also tasked to provide a category label for the security smell that it has described. We ask the LLM to make clusters of similar smell descriptions and label the similar security smells with the same category name. 

\item \textit{Security smell categorization via LLM mapping -}
From the previous prompt, the LLM performs an automatic mapping of a security category to the descriptions and labels every cluster of similar descriptions with a high-level security category name. In this example, ``Insecure Logging'' is linked to the broader category of ``Logging and Monitoring'' smells after we ask it to propose a categorization of the security smell. 
\end{itemize}

After conducting both CWE-based and LLM-based categorizations, we performed manual consistency analysis of the results. We found that LLM-generated categories consistently matched CWE-based categories. For example, the LLM-proposed category "Logging and Monitoring" corresponds to instances of "CWE-532: Inclusion of Sensitive Information in Log Files" and "CWE-217: Failure to Protect Critical Data when Storing Logs" in our CWE categories.

After manual mapping of both categorization methods' results, we observed a 95\% consistency rate between CWE categories and LLM-generated categories. This validates the sound capabilities of LLMs in generating correct security smell categories and extracting correct CVE details when provided with vulnerable IaC code snippets.

We chose to present LLM-proposed categories of security smells to enhance specificity and contextual relevance of security smell detection in IaC scripts. While CWE categories provide standardized, widely recognized labels, LLM-generated categories offer dynamically tailored descriptions that capture nuanced security risks directly tied to IaC configurations and patterns. This approach allows us to investigate the potential of these labels to improve detection precision and provide actionable insights beyond fixed CWE categorization limitations.

\subsection{RQ2: How recurrent and persistent are security smells in IaC scripts?}
\label{sec:smells}

We analyze the distribution and occurrences of identified security smell categories across IaC tools, then explore whether code changes addressing security smells are applied immediately after introduction. To assess the effectiveness of current security practices, we examine the lifespan of security smells in projects where the Top 10 security smells have been identified. First, we automatically investigate whether our identified security smells existed during prior taxonomy periods. Using labels from our Top 10 security smell categories, we mine GitHub repositories from our dataset to search for instances during the interval between our study and previous work. We manually validate findings and stop upon identifying at least one occurrence of each security smell.

We determine security smell recurrence using two attributes: occurrence number in our complete dataset of code changes and proportion in scripts. To study security smell evolution, we use the ``diff'' command line to automatically search for code changes associated with smelly code snippets in affected GitHub files through previous commits. We consider a smell persistent if we detect the same code snippet with identical variable names and values in the same script file across different project versions. We manually assess the number of commits required for maintainers to fix security smells. This experiment evaluates whether IaC maintainers address security smells immediately after insertion, their awareness of smelly code, and systematic use of security detection tools. Additionally, we confirm whether maintainers truly fix security smells, as prior work shows that some supposedly fixed script versions remain smelly~\cite{9321740}.

\subsection{RQ3: To what extent can security check rules derived from the enhanced taxonomy effectively improve existing IaC linters?}

We propose proactive mitigation strategies for IaC practitioners by designing, automatically generating, and validating security rules for static analysis of security smells in IaC scripts. We explore how practitioners can leverage our taxonomy to improve IaC script security by augmenting linter capabilities.

\subsubsection{Security Rules for the Static Analysis of Security Smells}
\label{sec:rules}
We design security rules to enhance smell detection capabilities for the ten most represented security smell categories in linters for our seven IaC tools: Ansible-lint for Ansible, Bandit and Yaml-Lint for Vagrant (the tools respectively covering both vagrantfiles and YAML configuration files from its dataset), Terrascan for Terraform, Rubocop for Puppet and Chef (both written in Ruby), ESLint for Pulumi, and Salt-Lint for SaltStack. 

In Table~\ref{tab:rules_design}, we manually design rules aimed at augmenting selected linters to detect the Top 10 security smells in IaC scripts. These rule designs are then used to automatically generate corresponding code for each linter with LLM assistance.

\begin{table}[!h]
    \caption{Security Rules for the Detection of the Top 10 Security Smells in IaC Scripts}
\label{tab:rules_design}

\scriptsize
  \resizebox{\linewidth}{!}{   \centering
\begin{tabular}{|p{.25\linewidth}|p{.8\linewidth}|}
\hline
\textbf{Security Smell} & \textbf{Rule Condition} \\ \hline

{Insecure Configuration Management} & 
\tt (isConfigFile(x) $\wedge$ isDefaultSetting(x) $\wedge$ isSensitiveSetting(x)) \\ \hline

{Insecure Dependency Management} & 
\tt (isDependency(x) $\wedge$ lacksVersionLocking(x) $\wedge$ isUntrustedSource(x)) \\ \hline

{Insecure Input Handling} & 
\tt (isInput(x) $\wedge$ lacksValidation(x) $\wedge$ isExploitable(x)) \\ \hline

{Outdated Dependencies} & 
\tt (isDependency(x) $\wedge$ isOutdatedVersion(x.version) $\wedge$ hasKnownVulnerabilities(x)) \\ \hline

{Path Traversal} & 
\tt (isFilePath(x) $\wedge$ isUserInput(x.value) $\wedge$ isUnsanitized(x.value)) \\ \hline

{Command Injection} & 
\tt (isUserInput(x) $\wedge$ isCommand(x) $\wedge$ isUnsanitized(x)) \\ \hline

{Code Injection} & 
\tt (isUserInput(x) $\wedge$ isCommand(x) $\wedge$ isUnsanitized(x)) \\ \hline

{Outdated Software Version} & 
\tt (isDependency(x) $\wedge$ isOutdatedVersion(x.version)) \\ \hline

{Inadequate Naming Conventions} & 
\tt (isVariable(x) $\wedge$ followsNonStandardConvention(x) $\wedge$ reducesReadability(x)) \\ \hline

{Sensitive Information Exposure} & 
\tt (isAttribute(x) $\wedge$ isSensitiveData(x.name) $\wedge$ isExposed(x)) \\ \hline

\end{tabular}

    }
\end{table}

\subsubsection{Security Rules Generation using LLMs}
To support automated rule generation for our selected IaC linters, we leverage an LLM to synthesize rule code. We compile a dataset of existing linter rules written in each tool's syntax, serving as demonstration material and reference for prompt engineering. The LLM is prompted with a structured template including the security rule design and example linter rule code, enabling generation of new rules tailored to each linter's framework. We select an LLM trained on datasets containing the programming languages associated with the IaC linters to increase output quality probability.

{\scriptsize
\begin{verbatim}
Prompt >>
    You are a code generation assistant specialized in
    writing rules for IaC linters. Given:
    - A security smell and its detection criteria.
    - Example rules written for other IaC linters. 
    Your task is to generate a detection rule for 
    Ansible-lint, Terrascan, Rubocop, ESLint, and Salt-lint
    that detects the provided security smell.
    
    Security Smell:
    "[SECURITY_SMELL]: [DETECTION_CRITERIA]"
    
    Example Rules:
    # Ansible-Lint
    [ANSIBLE_LINT_RULE_EXAMPLE]
    
    # Terrascan
    [TERRASCAN_RULE_EXAMPLE]
    
    # Rubocop
    [RUBOCOP_RULE_EXAMPLE]
    
    # ESLint
    [ESLINT_RULE_EXAMPLE]
    
    # Salt-Lint
    [SALT_LINT_RULE_EXAMPLE]
    
    Now, write the equivalent rule for each of the selected 
    linters that detects the provided security smell.
>>>
\end{verbatim}
}

\subsubsection{Test and Evaluation of Rules Efficiency}
We constructed an oracle dataset from the remaining code snippets in our initial corpus (remaining entries after construction of the seven datasets). From this, we manually sampled 212 representative smelly code snippets, covering the Top 10 security smells across all seven IaC tools. We then applied our augmented linters, enriched with LLM-generated rules that were manually curated and validated, to this oracle dataset. Finally, both authors and three independent external raters conducted an additional manual validation of the detection results. We evaluated the effectiveness of our rules using precision, defined as the proportion of true positives among all detections. Precision measures how often a rule correctly identifies a security smell when it detects one. We evaluated the precision of the LLM-generated rules that were manually validated on the Ansible, Terraform, Puppet, and Saltstack datasets. Additionally, we conducted an ablation study by reporting the precision of the raw LLM-generated rules without manual validation on our Vagrant, Chef, and Pulumi datasets. Together, these analyses provide a comprehensive assessment of the reliability and effectiveness of our LLM-based rule generation approach and demonstrate LLMs' capability to automate time-consuming tasks. All data are available in our replication package.

\section{Setup for LLM-Assisted Analysis with Human Validation}

To assist—not replace—human analysis in processing 1,094 code snippets across seven IaC tools, we employed a carefully controlled LLM pipeline with systematic human validation at each stage. We used \texttt{GPT-3.5-turbo-1106} and \texttt{GPT-4o} as initial processors, with all outputs subject to manual review, reconciliation, and validation. The models served complementary assistive roles: efficiency in structured extraction and initial pattern recognition in code analysis as well efficacy in rule generation.

\textbf{Extractive categorization.} \texttt{GPT-3.5-turbo-1106} received structured inputs consisting of the vulnerable code, its fixed version, and the corresponding commit message. The model’s task was to identify potential CVEs, then map them to relevant CWE categories. This division of labor leveraged the model’s efficiency and stability for structured extraction. In our work, we'll refer to GPT-3.5-turbo-1106 as simply GPT-3.5-turbo since it is the default model for that alias.

\textbf{Generative categorization.} \texttt{GPT-4o} analyzed pairs of vulnerable and fixed code snippets to generate detailed descriptions of security smells and propose high-level category labels by clustering similar descriptions. This model was chosen based on recent empirical evidence showing superior performance in vulnerability detection and structured code review tasks~\cite{bae2024gpt4o}. To avoid bias, commit messages were excluded from the generative inputs, ensuring categorizations relied solely on code semantics.
Furthermore, we used the model to assist us in enhancing linters associated with our seven types of IaC scripts, namely Ansible-Lint, Rubocop, Terrascan, Bandit, Yaml-Lint, Salt-Lint, and ESLint to detect our categorized security smells. Specifically, we leverage GPT-4o to automate the generation of static analysis rules that we manually correct and validate using them on their associated linters.

\textbf{Reproducibility and controlled prompting.} To maximize consistency in LLM assistance, we used deterministic parameters (temperature=0.1, top-p=1.0, maximum token limit=1000). These settings follow OpenAI’s recommended best practices for reproducible code generation~\cite{openai2023params}. Our prompting strategy combined structured task instructions with a few-shot design informed by validated domain examples, which guided models toward precise, functional outputs while reducing false positives.

However, we acknowledge that LLM outputs are inherently non-deterministic. Therefore, reproducibility of our work depends not on replicating exact LLM outputs, but on following our documented validation protocol. All prompts, validation criteria, intermediate results, and human decision logs are provided in our replication package, enabling others to scrutinize our quality control process.

\textbf{Multi-layered validation framework.}  Recognizing the critical importance of human oversight when using LLMs for foundational research tasks, we implemented a comprehensive validation protocol: \ding{172}\textbf{Stage 1 - Research team validation:} All LLM outputs underwent initial review by the authors, who corrected errors, resolved ambiguities, and ensured alignment with security best practices. \ding{173} \textbf{Stage 2 - Cross-reference with standards:} Every category was manually verified against CWE classifications and OWASP guidelines to ensure industry alignment.
\ding{174}  \textbf{Stage 3 - External expert review:} Three independent experts (two Master's students in cybersecurity and one PhD student in Software Engineering) evaluated a stratified sample of categorizations, mappings, and rules. They assessed correctness on a binary scale (1 for correct, 0 for incorrect) with detailed justifications. This review process aligns with established best practices for inter-rater reliability in software engineering research~\cite{gonzalez2020interrater}.
\ding{175}  \textbf{Stage 4 - Reconciliation:} Disagreements between LLM suggestions and human validators were resolved through discussion, with human judgment taking precedence in all cases.

\section{Study Results}
\label{sec:results}
We present the findings related to the research questions of our investigation in this section.

\subsection{Answer to RQ1: What security smells exist within IaC scripts?}

Our pipeline identified 472 security smells (CVEs) across 62 CWE-related categories in IaC scripts.
Figure~\ref{fig:ansible_playbook_real_world} presents the 10 most represented security smell categories, which are as follows:
\begin{figure}[!htp]
    \centering
    \begin{adjustbox}{max width=.9\linewidth}
    \begin{lstlisting}[language=yaml, numbers=left]
- name: Example Playbook with Real-World Security Smells
  hosts: all
  tasks:
    - name: Misconfigured SSH service
      lineinfile:
        path: /etc/ssh/sshd_config
        regexp: '^PermitRootLogin'
        line: 'PermitRootLogin yes' #1 - Insecure Configuration Management

    - name: Install a package without checking version
      apt:
        name: "apache2" #2 - Insecure Dependency Management
        state: present

    - name: Process user-supplied number without validation
      shell: "echo $(( {{ user_number }} + 1 ))" #3 - Insecure Input Handling

    - name: Install package with outdated dependencies
      apt:
        name: "openssl"
        version: "1.0.1" #4 - Outdated Dependencies
        state: present

    - name: Copy file with potential path traversal
      copy:
        src: "{{ file_path }}" #5 - Path Traversal
        dest: /etc/securefile

    - name: Run package update command from input
      command: "apt-get {{ action }}" #6 - Command Injection

    - name: Evaluate user-provided Python expression (vulnerable)
      vars:
        user_expression: "os.system('rm -rf /')"
      set_fact:
        result: "{{ lookup('pipe', 'python3 -c \"' + user_expression + '\"') }}" #7 - Code Injection

    - name: Install outdated version of Python
      apt:
        name: "python2.7" #8 - Outdated Software Version 
        state: present

    - name: Create file with vague name
      file:
        path: /etc/doitnow.txt
        state: touch #9 - Inadequate Naming Convention

    - name: Store AWS credentials in plain text file
      copy:
        content: |
          [default]
          aws_access_key_id = {{ aws_access_key_id }}
          aws_secret_access_key = {{ aws_secret_access_key }} 
        dest: /etc/aws/credentials
        owner: root
        mode: '0600' #10 - Sensitive Information Exposure
\end{lstlisting}
    \end{adjustbox}
    \caption{Ansible Playbook with Real-World Examples of our Top 10 Security Smells}
    \label{fig:ansible_playbook_real_world}
\end{figure}

\begin{itemize}[noitemsep,topsep=0pt,left=0pt]
\item[\wcircle{1}] \textit{Insecure Configuration Management}: Allowing root login over SSH enables unauthorized access (CWE-306: Missing Authentication for Critical Function).
\item[\wcircle{2}] \textit{Insecure Dependency Management}: Installing packages without version constraints may introduce vulnerable versions (CWE-1104: Use of Unmaintained Third Party Components).

\item[\wcircle{3}] \textit{Insecure Input Handling}: Processing user input without validation in shell commands (CWE-20: Improper Input Validation).

\item[\wcircle{4}] \textit{Outdated Dependencies}: Using packages with known vulnerabilities, such as OpenSSL 1.0.1 (CWE-1104: Use of Unmaintained Third Party Components).

\item[\wcircle{5}] \textit{Path Traversal}: Using unvalidated user-supplied file paths allows unauthorized file access (CWE-22: Improper Limitation of a Pathname to a Restricted Directory).

\item[\wcircle{6}] \textit{Command Injection}: Incorporating unsanitized user input into system commands (CWE-77: Command Injection).

\item[\wcircle{7}] \textit{Code Injection}: Executing commands based on unvalidated user input (CWE-94: Improper Control of Generation of Code). The input is interpreted as Python code in this example, not just a shell command. Attackers could execute arbitrary Python instructions.

\item[\wcircle{8}] \textit{Outdated Software Version}: Installing deprecated software like Python 2.7 (CWE-1104: Use of Unmaintained Third Party Components).

\item[\wcircle{9}] \textit{Inadequate Naming Convention}: Non-descriptive resource naming hinders maintainability and security audits (CWE-710: Improper Adherence to Coding Standards).

\item[\wcircle{10}] \textit{Sensitive Information Exposure}: Storing credentials in plain text files (CWE-256: Unprotected Storage of Credentials).
\end{itemize}

Each of the presented smells poses a security risk in IaC scripts like the examples provided in the Ansible playbook (Figure~\ref{fig:ansible_playbook_real_world}). By failing to sanitize input, improperly handling sensitive data, or allowing dangerous user-controlled data in commands, practitioners can open up their systems to various attack vectors. 
Our findings connect well with the security smells studied in the literature and overlap with the CWE categories from our CWE-based taxonomy of smells, therefore demonstrating the consistency of our results, as well as underscoring the importance of addressing and mitigating these issues. 

Overall, the detected categories of security smells in our work align with the OWASP Top 10 CI/CD of security risks\footnote{\url{https://owasp.org/www-project-top-10-ci-cd-security-risks}} released in 2022. This initiative of providing a list of security risks for CI/CD also stresses the importance of sensitization and mitigation of these issues for practitioners and industries.

\finding{
\footnotesize
{\bf Insights from RQ1: \ding{224}} 
Our results show that IaC scripts exhibit widespread security weaknesses, with 472 smells across 62 CWE-related categories, mirroring the OWASP Top 10 CI/CD risks. 
Automated detection can aid practitioners, but lasting improvements in IaC security require balancing automation with expert oversight.

}

\subsection{Answer to RQ2: How recurrent and persistent are security smells in IaC scripts?}

\subsubsection{Distribution of Security Smells by Script Type}

We analyze the distribution and frequency of the Top 10 security smell categories across our seven selected IaC tools through Figures~\ref{fig:distribution} and~\ref{fig:persistence}. Figure~\ref{fig:distribution} shows the relative percentage of each smell per tool, while Figure~\ref{fig:persistence} illustrates the temporal evolution of security smells from 2019-2024.

\newcommand{\legenditem}[2]{%
    \tikz[baseline=-0.6ex] \node[draw=black, fill=#1, circle, minimum size=5pt, inner sep=0pt] {};~#2%
}

\begin{figure}[H]
\centering
\scriptsize
\begin{adjustbox}{width=.8\linewidth}
\begin{minipage}[t]{.3\linewidth}
    \centering
    \begin{adjustbox}{width=\textwidth}
    \centering
    \begin{tikzpicture}
    \begin{axis}[
    ybar stacked,
    symbolic x coords={Ansible,Terraform,Chef,Puppet,Pulumi,Vagrant,Saltstack},
    xtick=data,
    ymin=0, ymax=100,
    ylabel={Percentage of Security Smells},
    x tick label style={rotate=90, anchor=east, font=\footnotesize},width=4cm,
height=4.5cm,
bar width=3pt,
            enlarge x limits=0.2,
]

\addplot+[fill=gray!10] coordinates {
    (Ansible,  0.00) (Terraform, 4.37) (Chef, 74.90)
    (Puppet,  0.11) (Pulumi, 0.00) (Vagrant, 3.80)
    (Saltstack, 0.00)
};

\addplot+[fill=gray!30] coordinates {
    (Ansible, 37.67) (Terraform,10.84) (Chef, 3.35)
    (Puppet,  3.54) (Pulumi, 2.68) (Vagrant, 3.80)
    (Saltstack, 5.68)
};

\addplot+[fill=gray!50] coordinates {
    (Ansible,  6.68) (Terraform, 8.21) (Chef, 3.38)
    (Puppet,  5.31) (Pulumi, 2.21) (Vagrant,16.79)
    (Saltstack,22.58)
};

\addplot+[fill=black!50] coordinates {
    (Ansible,  0.95) (Terraform,13.30) (Chef, 1.85)
    (Puppet,  0.35) (Pulumi, 3.62) (Vagrant,59.33)
    (Saltstack,20.56)
};

\addplot+[fill=white] coordinates {
    (Ansible,  4.21) (Terraform, 1.03) (Chef, 0.12)
    (Puppet,  0.10) (Pulumi, 2.71) (Vagrant, 0.11)
    (Saltstack, 0.00)
};

\addplot+[fill=brown!30] coordinates {
    (Ansible,  3.17) (Terraform,10.22) (Chef, 3.16)
    (Puppet, 55.89) (Pulumi,18.18) (Vagrant, 3.64)
    (Saltstack, 0.00)
};

\addplot+[fill=brown!60] coordinates {
    (Ansible,  3.15) (Terraform,28.47) (Chef, 0.00)
    (Puppet, 13.98) (Pulumi,17.41) (Vagrant, 3.64)
    (Saltstack, 0.00)
};

\addplot+[fill=brown!80] coordinates {
    (Ansible, 17.67) (Terraform,10.76) (Chef, 4.57)
    (Puppet, 15.96) (Pulumi, 3.35) (Vagrant, 1.07)
    (Saltstack,12.33)
};

\addplot+[fill=orange!40] coordinates {
    (Ansible, 23.58) (Terraform,12.08) (Chef, 4.12)
    (Puppet,  1.12) (Pulumi,44.25) (Vagrant, 5.22)
    (Saltstack,17.07)
};

\addplot+[fill=orange!70] coordinates {
    (Ansible,  2.93) (Terraform, 0.74) (Chef, 4.58)
    (Puppet,  3.63) (Pulumi, 5.62) (Vagrant, 2.65)
    (Saltstack,21.82)
};

\end{axis}
\end{tikzpicture}
\end{adjustbox}
  \captionsetup{font=footnotesize}
\caption{Distribution of Security Smells in Different IaC Scripts}
\label{fig:distribution}
\end{minipage}

\hspace{0.05\textwidth}

\begin{minipage}[t]{.35\linewidth}
    \centering
    \begin{adjustbox}{width=\textwidth}
    \centering
    \begin{tikzpicture}
\begin{axis}[
    ybar stacked,
            symbolic x coords={2019,2020,2021,2022,2023,2024},
            xtick=data,
            ymin=0, ymax=100,
            ylabel={Percentage of Security Smells},
            xtick style={draw=none},
            x tick label style={rotate=90, anchor=east, font=\footnotesize},
            width=4cm,
height=4.5cm,
bar width=3pt,
enlarge x limits=0.1,
]

\addplot+[fill=gray!10] coordinates {(2019,4.66) (2020,14.09) (2021,7.57) (2022,0.06) (2023,0.13) (2024,24.21)}; 
\addplot+[fill=gray!30] coordinates {(2019,28.63) (2020,8.61) (2021,4.72) (2022,20.10) (2023,21.10) (2024,1.69)}; 
\addplot+[fill=gray!50] coordinates {(2019,2.90) (2020,14.89) (2021,31.88) (2022,7.02) (2023,20.20) (2024,8.41)}; 

\addplot+[fill=black!50] coordinates {(2019,0.40) (2020,23.55) (2021,8.29) (2022,11.27) (2023,13.49) (2024,10.59)}; 
\addplot+[fill=white] coordinates {(2019,5.98) (2020,4.50) (2021,0.68) (2022,0.51) (2023,2.33) (2024,1.13)};   
\addplot+[fill=brown!30] coordinates {(2019,24.55) (2020,5.61) (2021,16.20) (2022,9.06) (2023,8.98) (2024,8.59)}; 

\addplot+[fill=brown!60] coordinates {(2019,12.65) (2020,0.69) (2021,0.67) (2022,3.81) (2023,5.00) (2024,23.24)};  
\addplot+[fill=brown!80] coordinates {(2019,18.92) (2020,13.59) (2021,11.33) (2022,20.62) (2023,16.21) (2024,2.64)}; 
\addplot+[fill=orange!40] coordinates {(2019,1.11) (2020,1.21) (2021,0.93) (2022,15.51) (2023,8.75) (2024,16.53)};   
\addplot+[fill=orange!70] coordinates {(2019,0.21) (2020,13.24) (2021,17.73) (2022,12.05) (2023,3.81) (2024,2.97)};  

\end{axis}
\end{tikzpicture}
\end{adjustbox}
\captionsetup{font=footnotesize}
\caption{Persistence of Security Smells (2019–2024)}
\label{fig:persistence}
\end{minipage}

\end{adjustbox}
\scriptsize
\captionsetup[table]{justification=raggedright,singlelinecheck=false}
    \begin{tabular}{cc}
        \legenditem{gray!10}{Outdated Software Version} &
        \legenditem{gray!30}{Insecure Configuration Management} \\
        \legenditem{gray!50}{Outdated Dependencies} &
        \legenditem{black!50}{Path Traversal} \\
        \legenditem{white}{Sensitive Information Exposure} &
        \legenditem{brown!30}{Code Injection} \\
        \legenditem{brown!60}{Command Injection} &
        \legenditem{brown!80}{Insecure Input Handling} \\
        \legenditem{orange!40}{Insecure Dependency Management} &
        \legenditem{orange!70}{Inadequate Naming Convention}  \\
    \end{tabular}
\end{figure}

\noindent
\textbf{Tool-Specific Security Patterns}
Our analysis reveals distinct security smell profiles across IaC tools. \textit{Chef} shows exceptional prevalence of \textit{Outdated Software Version} smells (74.9\%), indicating systemic version management issues. \textit{Vagrant} and \textit{Saltstack} exhibit high \textit{Path Traversal} rates (59.33\% and 20.56\%), suggesting inadequate path sanitization practices. \textit{Puppet} demonstrates significant \textit{Code Injection} vulnerability (55.89\%), likely from embedded shell commands and templating mechanisms.

\noindent
\textbf{Cross-Tool Vulnerabilities}
\textit{Insecure Dependency Management} and \textit{Insecure Input Handling} affect multiple tools, with \textit{Pulumi} (44.25\%) and \textit{Saltstack} (22.58\%) showing elevated rates. These patterns indicate common challenges in third-party module management and input validation across cloud automation platforms.

\noindent
\textbf{Low-Frequency, High-Impact Issues}
\textit{Sensitive Information Exposure} remains rare (typically $\leq$1\%) but poses significant security risks when present. \textit{Inadequate Naming Convention} appears prominently in \textit{Saltstack} (21.82\%), affecting maintainability though with limited direct security impact.

\noindent
\textbf{Implications.}
The skewed distribution highlights how community practices and tool architectures shape security postures. Results indicate that tool-specific guidelines, enhanced static analysis capabilities, and targeted developer training are essential for addressing recurring vulnerabilities and securing infrastructure automation practices.

\subsubsection{Persistence of Security Smells in IaC Scripts}

\begin{figure}[H]
    \centering
    \begin{adjustbox}{max width=\linewidth}
    \begin{tikzpicture}
        \begin{groupplot}[
            group style={
                group size=5 by 2,
                horizontal sep=1.5cm,
                vertical sep=2cm
            },
            width=0.35\textwidth,
            height=3cm,
            ymin=0, ymax=80,
            ybar,
            nodes near coords,
            every node near coord/.append style={font=\footnotesize, anchor=south},
            symbolic x coords={2019, 2020, 2021, 2022, 2023, 2024},
            xtick=data,
            xtick style={draw=none},
            x tick label style={rotate=45, anchor=east, font=\footnotesize},
            legend style={at={(0.5,-0.15)}, anchor=north, legend columns=-1},
            cycle list name=color list,
            enlarge x limits=0.15
        ]

        \pgfplotscreateplotcyclelist{color list}{
            {blue!50},
            {green!50},
            {red!50},
            {orange!50},
            {purple!50},
            {cyan!50},
            {magenta!50},
            {gray!50},
            {yellow!50},
            {teal!50},
            {brown!50}
        }

        \nextgroupplot[title={(a) Code Injection}]
        \addplot+[fill=brown!30] coordinates {(2019,39) (2020,11) (2021,35) (2022,23) (2023,26) (2024,26)};

        \nextgroupplot[title={(b) Command Injection}]
        \addplot+[fill=brown!60] coordinates {(2019,20) (2020,1) (2021,1) (2022,10) (2023,14) (2024,71)};

        \nextgroupplot[title={(c) Inadequate Naming Convention}]
        \addplot+[fill=orange!70] coordinates {(2019,0) (2020,26) (2021,38) (2022,31) (2023,11) (2024,9)};

        \nextgroupplot[title={(d) Insecure Configuration Management}]
        \addplot+[fill=gray!30] coordinates {(2019,46) (2020,17) (2021,10) (2022,51) (2023,61) (2024,5)};

        \nextgroupplot[title={(e) Insecure Dependency Management}]
        \addplot+[fill=orange!40] coordinates {(2019,1) (2020,2) (2021,2) (2022,40) (2023,25) (2024,50)};

        \nextgroupplot[title={(f) Insecure Input Handling}]
        \addplot+[fill=brown!80] coordinates {(2019,30) (2020,27) (2021,24) (2022,53) (2023,47) (2024,8)};

        \nextgroupplot[title={(g) Outdated Dependencies}]
        \addplot+[fill=gray!50] coordinates {(2019,4) (2020,29) (2021,69) (2022,18) (2023,58) (2024,26)};

        \nextgroupplot[title={(h) Sensitive Information Exposure}]
        \addplot+[fill=white] coordinates {(2019,9) (2020,9) (2021,1) (2022,1) (2023,7) (2024,3)};

        \nextgroupplot[title={(i) Path Traversal}]
        \addplot+[fill=black!50] coordinates {(2019,0) (2020,47) (2021,18) (2022,29) (2023,39) (2024,32)};

        \nextgroupplot[title={(j) Outdated Software Version}]
        \addplot+[fill=gray!10] coordinates {(2019,7) (2020,28) (2021,16) (2022,0) (2023,0) (2024,74)};

        \end{groupplot}
    \end{tikzpicture}
    \end{adjustbox}
  \captionsetup{font=footnotesize}
    \caption{Persistence of Security Smells in Detail (2019-2024)}
    \label{fig:detailled_persistence}
\end{figure}
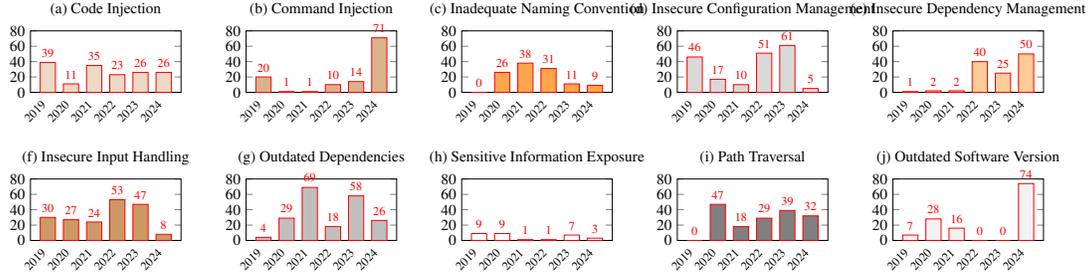

We analyze the temporal evolution of security smells through Figures~\ref{fig:persistence} and~\ref{fig:detailled_persistence}, revealing how the Top 10 categories persist, emerge, or decline over the five-year period from 2019 to 2024.

\noindent
\textbf{Early Prevalence and Mid-Period Shifts}
Between 2019 and 2021, \textit{Insecure Configuration Management} exhibits dramatic decline (28.6\% to 4.7\%) while \textit{Outdated Dependencies} surges (2.9\% to 31.9\%). This reversal suggests initial focus on configuration hardening followed by inadequate dependency vetting as projects rapidly adopted new libraries.

\noindent
\textbf{Persistent Injection Risks}
Both injection types remain consistently problematic. \textit{Code Injection} maintains 16-24\% prevalence annually, while \textit{Command Injection} spikes from 0.7\% (2020) to 23.2\% (2024), with raw counts jumping from single-digits to 71 occurrences. These trends underscore persistent challenges in sanitizing dynamic inputs in IaC scripts.

\noindent
\textbf{Emerging Path and Dependency Concerns}
\textit{Path Traversal} grows steadily from 0.4\% (2019) to 10.6\% (2024), with raw counts rising from 0 to 32 occurrences. Similarly, \textit{Insecure Dependency Management} climbs from 1.1\% to 16.5\%, reflecting a shift from static configuration issues to dynamic package ecosystem vulnerabilities.

\noindent
\textbf{Late-Period Anomalies}
\textit{Outdated Software Version} shows the most pronounced anomaly, surging to 24.2\% in 2024 (74 occurrences) after remaining below 8\% through 2023. This reversal likely stems from legacy modules being used without automated update mechanisms.

\noindent
\textbf{Stable Low-Frequency Smells}
\textit{Sensitive Information Exposure} and \textit{Inadequate Naming Convention} remain minority concerns (under 3\% each), though raw counts indicate continued relevance for maintainability and confidentiality.

\noindent
\textbf{Adaptive Mitigation Strategies.}
While some smells decline as best practices emerge, others persist or resurge, demonstrating that IaC security debt evolves with tooling ecosystems. The growing adoption of IaC tools (evidenced by package download statistics from npm, PyPI, and Stack Overflow Developer Surveys 2019-2024) exacerbates these issues as expanding communities introduce inconsistent security practices. The recurrence pattern indicates practitioners continue making similar security mistakes, reinforcing the need for continuously updated mitigation strategies combining static checks, runtime validation, and automated dependency governance.

\finding{
\footnotesize
{\bf Insights from RQ2: \ding{224}}
Security smells exhibit distinct distribution patterns across IaC tools: Chef scripts show 74.9\% outdated software issues while Puppet demonstrates 55.89\% code injection vulnerabilities. Despite growing industry adoption, these smells persist throughout 2019-2024, indicating that IaC development lacks automated detection and remediation practices, causing practitioners to repeat the same security mistakes as tooling evolves.
}

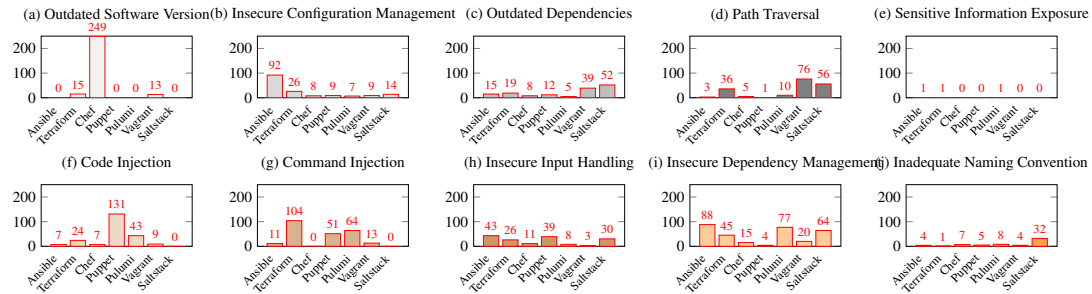
\begin{figure}[H]
    \centering
    \begin{adjustbox}{max width=\linewidth}
    \begin{tikzpicture}
        \begin{groupplot}[
            group style={
                group size=5 by 2, 
                horizontal sep=1.5cm,
                vertical sep=2cm
            },
            width=0.35\textwidth,
            height=3cm,
            ymin=0, ymax=250, 
            ybar,
            nodes near coords,
every node near coord/.append style={font=\footnotesize, rotate=0, anchor=south},
            symbolic x coords={Ansible, Terraform, Chef, Puppet, Pulumi, Vagrant, Saltstack},
            xtick=data,
            xtick style={draw=none}, 
            x tick label style={rotate=45, anchor=east, font=\footnotesize}, 
            legend style={at={(0.5,-0.15)}, anchor=south, legend columns=-1},
            cycle list name=color list,
            enlarge x limits=0.15
        ]

        \pgfplotscreateplotcyclelist{color list}{
            {blue!50},
            {green!50},
            {red!50},
            {orange!50},
            {purple!50},
            {cyan!50},
            {magenta!50},
            {gray!50},
            {yellow!50},
            {teal!50}
        }

    \nextgroupplot[title={(a) Outdated Software Version}]
    \addplot+[fill=gray!10] coordinates {
      (Ansible,   0) (Terraform,  15) (Chef, 249)
      (Puppet,    0) (Pulumi,      0) (Vagrant, 13)
      (Saltstack, 0)
    };

    \nextgroupplot[title={(b) Insecure Configuration Management}]
    \addplot+[fill=gray!30] coordinates {
      (Ansible,  92) (Terraform, 26) (Chef,  8)
      (Puppet,    9) (Pulumi,    7) (Vagrant,  9)
      (Saltstack,14)
    };

    \nextgroupplot[title={(c) Outdated Dependencies}]
    \addplot+[fill=gray!50] coordinates {
      (Ansible,  15) (Terraform,19) (Chef,  8)
      (Puppet,   12) (Pulumi,    5) (Vagrant,39)
      (Saltstack,52)
    };

    \nextgroupplot[title={(d) Path Traversal}]
    \addplot+[fill=black!50] coordinates {
      (Ansible,   3) (Terraform,36) (Chef,   5)
      (Puppet,    1) (Pulumi,   10) (Vagrant,76)  
      (Saltstack,56)
    };

    \nextgroupplot[title={(e) Sensitive Information Exposure}]
    \addplot+[fill=white] coordinates {
      (Ansible,  1) (Terraform, 1) (Chef, 0)
      (Puppet,   0) (Pulumi,  1) (Vagrant, 0)
      (Saltstack,0)
    };

    \nextgroupplot[title={(f) Code Injection}]
    \addplot+[fill=brown!30] coordinates {
      (Ansible,  7) (Terraform,24) (Chef,  7)
      (Puppet, 131) (Pulumi, 43) (Vagrant, 9)
      (Saltstack,0)
    };

    \nextgroupplot[title={(g) Command Injection}]
    \addplot+[fill=brown!60] coordinates {
      (Ansible, 11) (Terraform,104) (Chef,  0)
      (Puppet,  51) (Pulumi, 64) (Vagrant,13)
      (Saltstack,0)
    };

    \nextgroupplot[title={(h) Insecure Input Handling}]
    \addplot+[fill=brown!80] coordinates {
      (Ansible, 43) (Terraform,26) (Chef, 11)
      (Puppet,  39) (Pulumi,  8) (Vagrant, 3)
      (Saltstack,30)
    };

    \nextgroupplot[title={(i) Insecure Dependency Management}]
    \addplot+[fill=orange!40] coordinates {
      (Ansible, 88) (Terraform,45) (Chef, 15)
      (Puppet,   4) (Pulumi,  77) (Vagrant,20)  
      (Saltstack,64)
    };

    \nextgroupplot[title={(j) Inadequate Naming Convention}]
    \addplot+[fill=orange!70] coordinates {
      (Ansible,  4) (Terraform, 1) (Chef,  7)
      (Puppet,    5) (Pulumi,  8) (Vagrant, 4)
      (Saltstack,32)
    };

        \end{groupplot}
    \end{tikzpicture}
    \end{adjustbox}
  \captionsetup{font=footnotesize}
    \caption{Detailled Distribution of Security Smells in Different Infrastructure as Code Scripts}
    \label{fig:detailled_distribution}
\end{figure}

\subsection{Answer to RQ3: To what extent can security check rules derived from the enhanced taxonomy effectively improve existing IaC linters?}
Our third research question aimed to assess the usefulness of our taxonomy of security smells in the enhancement of IaC scripts' security. We have automatically generated rules using an LLM, namely GPT-4o, to augment the capabilities of four linters (Ansible-Lint, Terrascan, Rubocop, ESLint, and Salt-Lint) for the detection of our 62 categories of security smells. However, we have only manually curated and validated so far the rules in our Top 10 categories of security smells. We first automatically executed the augmented linters on our Oracle dataset, which consists of 212 smelly IaC code snippets covering samples from our Top 10 security smell categories. Then, we manually corrected and validated the warnings produced by the linters. We measure the efficiency of the linters in detecting the Top 10 using the precision metric. We compute it taking into account our manual corrections and the evaluation from our three raters. All data are available in our replication package. From the results reported with a perfect score, we observe that our taxonomy of security smells can also have a practical impact on practitioners and help to build safer IaC scripts through proactive and automated smell detection.

\begin{table}[h!]
\centering
\caption{Precision of LLM-generated Security Rules Based on External Ratings}
\label{tab:manually_validated_rules}
\resizebox{\textwidth}{!}{%
\begin{tabular}{@{}lc cccc|ccc@{}}
\toprule
\multirow{2}{*}{\textbf{Top 10 Categories of Security Smells}} & \multirow{2}{*}{\textbf{Occurrences}} & \multicolumn{4}{c|}{\textbf{Precision of Manually Validated Rules}} & \multicolumn{3}{c}{\textbf{Precision of LLMs Validated Rules}} \\
\cmidrule{3-6} \cmidrule{7-9}
& & \textbf{Ansible-lint} & \textbf{Rubocop - Puppet} & \textbf{Salt-Lint} & \textbf{Terrascan} & \textbf{Rubocop - Chef} & \textbf{Bandit + Yaml-Lint} & \textbf{ESLint} \\
\midrule
Command Injection              & 14 & 1.00 & --   & 1.00 & -- & --   & 0.25 & -- \\
Path Traversal                 & 23 & 1.00 & --   & 1.00 & -- & --   & 0.33 & 0.67 \\
Insecure Dependency Management &  5 & 1.00 & --   & --   & -- & --   & --   & --   \\
Insecure Configuration Management    & 60 & 1.00 & 1.00 & 1.00 & 0  & 0    & 0.42 & 0.44 \\
Insecure Input Handling        & 41 & 1.00 & 1.00 & 1.00 & -- & 0.36 & 0    & 0.25 \\
Code Injection                 &  7 & 1.00 & --   & 1.00 & -- & --   & --   & 1.00 \\
Outdated Software Dependencies &  1 & 1.00 & --   & --   & -- & --   & --   & --  \\
Outdated Software Version      &  3 & --   & --   & --   & 0  & --   & --   & --   \\
Sensitive Information Exposure & 58 & 1.00 & 1.00 & 1.00 & 0  & 0.25 & 0.25 & 0.60 \\
Inadequate Naming Convention   &  0 & --   & --   & --   & -- & --   & --   & --     \\
\midrule
\textbf{Total and AVG Precision}          & \textbf{212} & \textbf{1.00} & \textbf{1.00} & \textbf{1.00} & \textbf{0} & \textbf{0.20} & \textbf{0.25} & \textbf{0.59}  \\
\bottomrule
\end{tabular}%
}
\end{table}

In Table~\ref{tab:manually_validated_rules}, the results highlight the effect of removing manual validation in what can be seen as an ablation experiment. When LLM-generated rules are validated with the aid of existing static analyzers (e.g., Ansible-Lint, Rubocop for Puppet, Salt-Lint), precision remains consistently high, often reaching 1.00 across categories. In contrast, when validation relies solely on other LLM-based tools, precision drops substantially, with averages ranging from 0.20 (Rubocop for Chef) to 0.59 (ESLint). This gap is especially visible in high-frequency categories such as Insecure Configuration Management and Sensitive Information Exposure, where manual validation maintains high precision but LLM-only validation produces inconsistent results. These findings indicate that, while LLMs can generate useful rules, human-in-the-loop validation is essential to ensure reliability, as removing it leads to a significant reduction in precision. We consistently perform manual validation and rely on our external raters' decisions to validate our work for the enhancement of our selected linters.
However, in the case of Terrascan for Terraform, the results are particularly poor because the LLM often produced rules that were not compilable in the tool’s native format. Unlike with other linters, where generated rules could be manually adjusted and validated, Terrascan’s rule specification made this process extremely difficult: even small deviations from the expected syntax prevented execution, leaving us unable to effectively validate or refine the generated rules. Consequently, the observed precision for Terrascan does not only reflect the quality of the LLM output, but also the challenges of manual validation in this specific environment. This suggests that certain tools impose higher barriers to human-in-the-loop validation, which amplifies the gap between generated rules and their practical usability.

\finding{
{\bf Insights from RQ3: \ding{224}}
The key takeaway from our results is that LLMs alone are not enough. While they can rapidly generate IaC security rules from our taxonomy, their accuracy drops significantly without manual validation. With human-in-the-loop validation, precision reaches consistently high levels, but when validation relies solely on LLMs, results degrade sharply. This ablation experiment demonstrates that our taxonomy provides a practical foundation for enhancing linters, but also that manual curation remains indispensable for producing reliable, compilable, and actionable rules. Our taxonomy provides the structured foundation to make this human–AI collaboration effective. The Terrascan case further illustrates that certain tool ecosystems introduce additional barriers to rule validation, exacerbating this gap in LLM usage.
}

\section{Discussion}
\label{sec:discussion}
We discuss in this section the potential of LLMs into automating research work to some extent as well as the limitations and threats to validity inherent in our work.

\subsection{LLMs as Research Assistants: Opportunities and Responsibilities}

While this study demonstrates that LLMs can assist in accelerating certain research workflows, our experience underscores that they must be employed as tools under systematic human supervision rather than autonomous agents. In our approach to categorizing security smells and generating static analysis rules, LLMs served as initial processors that required extensive human validation at each stage. Specifically, we implemented a multi-layered quality assurance framework: (1) dual taxonomy derivation comparing LLM-generated categories against CWE standards, (2) manual reconciliation of all categorizations by multiple researchers, (3) external annotator validation, and (4) iterative refinement of generated rules through expert review. This human-in-the-loop approach was essential—without it, the taxonomy would lack the conceptual rigor necessary for a foundational contribution. Our experience suggests that while LLMs can reduce initial manual effort in pattern recognition tasks, they cannot replace the critical thinking and domain expertise required for establishing scientific taxonomies.
Nevertheless, our experience also hints that, beyond the scope of static analysis, this method can be extended to streamline tasks such as conducting Systematic Literature Reviews (SLRs), synthesizing taxonomies, or mapping studies where pattern recognition and rule-based classification are essential. By reducing the manual overhead typically associated with such processes, LLM-based pipelines can nowadays accelerate exploratory and maintenance tasks in software engineering research.

\subsection{Limitations and Threats to Validity}

Overall, we have identified several limitations and threats to validity in our study.  

First and foremost, we acknowledge that delegating taxonomy construction—a task requiring deep security expertise and careful judgment—to LLMs without systematic human oversight would be methodologically unsound. Our use of LLMs was therefore strictly bounded: they served as initial processors to handle the scale of 1,094 code snippets, but every output was subject to human validation. The final taxonomy represents human reasoning applied to LLM-assisted clustering, not raw model outputs. Indeed, to ensure the taxonomy's validity despite LLM involvement, we implemented systematic quality controls:
\begin{itemize}
    \item \textit{Dual-source validation}: Every LLM-generated category was cross-referenced against established CWE classifications, with discrepancies manually resolved.
    \item \textit{Multi-stage human review}: Each clustering decision underwent review by at least two authors, with disagreements resolved through discussion.
    \item \textit{Iterative refinement}: Generated rules underwent multiple rounds of manual inspection and testing against known vulnerabilities.
    \item \textit{External validation}: Independent annotators evaluated a sample of categorizations to assess inter-rater reliability.
    \item \textit{Prompt stability testing}: We conducted sensitivity analyzes using varied prompts to ensure consistent categorization.
\end{itemize}

Second, our validation experiments focused on the Top 10 security smells, despite generating rules for all 62 categories. This choice allowed for tractable evaluation but leaves the coverage of the remaining smells to be validated in future iterations.  

Third, our taxonomy reproducibility is limited by the non-deterministic nature of LLMs. The specific outputs depend on the model version, provider, parameters, and prompt phrasing. We mitigate this by fully detailing them in our experimental setup and in the provided replication GitHub repository. As a result, replication of our taxonomy or rules' generation can be facilitated.  

Fourth, while some recent works have raised concerns about prompting being potentially harmful or unreliable~\cite{morris2024prompting,perez2022ignore}, our use of LLMs is deliberate, controlled, and task-specific. We employ system prompts primarily for automating well-defined and step-by-step tasks such as clustering CWE categories, generating initial smell descriptions, and mapping descriptions to categories. These tasks are bounded in scope and do not involve open-ended creative generation, thereby reducing risks of unsafe or spurious content. 
Importantly, we treat LLMs as tools for assisting with repetitive or large-scale processing rather than as substitutes for researcher judgment. Each LLM-produced output was systematically followed by manual validation and reconciliation by the authors, and in many cases also verified by independent external raters. We further applied structured system prompts to constrain outputs, reduce ambiguity, and improve consistency. This workflow allowed us to harness the scalability benefits of LLMs while maintaining rigor and accuracy in our taxonomy construction and rule generation.
Thus, prompting in our study serves as a safe and effective automation aid. The final taxonomy and rules reflect human reasoning and validation rather than raw model outputs, ensuring that our findings remain reproducible, interpretable, and aligned with established security taxonomies.

Fifth, another concern could be data leakage from pretraining. However, our evaluation dataset is highly domain-specific, making overlap with LLM pretraining less probable than code written in general-purpose programming languages. In addition, current LLMs are not explicitly optimized for fine-grained IaC vulnerabilities~\cite{li2025everything,yin2024multitask}, which reduces, but does not eliminate, the risk of memorization. To mitigate this, we focused our efforts to build a categorization consensus between our two taxonomy methods on CWEs and high-level LLM-generated smell categories. 

Finally, we have not yet investigated the perception of practitioners regarding the relevance and usefulness of the identified smells and rules. Although our categories align with CWE/OWASP guidance, their practical value and usability in real-world development remain to be validated. Future work will include developer studies and tool-integration experiments to evaluate workflow integration, false-positive rates, and adoption.  

\subsection{Methodological Lessons for LLM-Assisted Security Research}
Our experience yields important insights for researchers considering LLM integration in security-critical taxonomies:
\begin{enumerate}
    \item LLMs should augment, not replace, expert judgment in foundational tasks
    \item Multiple validation layers are essential when LLMs contribute to core contributions
    \item Transparency about LLM involvement and validation processes is crucial for reproducibility
    \item The non-deterministic nature of LLMs requires explicit documentation of prompts, parameters, and validation criteria
\end{enumerate}

\section{Conclusion}
\label{sec:conclusion}
This study presents a comprehensive taxonomy of security smells in Infrastructure as Code scripts, identifying persistent categories across seven major IaC tools through systematic analysis of vulnerable script excerpts. Our findings reveal critical tool-specific vulnerability patterns. Chef scripts show 74.9\% outdated software issues while Puppet demonstrates 55.89\% code injection vulnerabilities. This shows that these smells persist and evolve despite the growing industry adoption.

The persistence of these vulnerabilities exposes a fundamental gap: practitioners lack systematic detection and remediation practices, repeatedly making the same security mistakes as tooling evolves. Our LLM-assisted rule generation approach successfully enhances existing linters, providing immediate practical improvements, and demonstrating the viability of automated detection enhancement. 

As infrastructure automation becomes the backbone of digital transformation, security smells in IaC scripts represent vulnerabilities that scale across entire Cloud ecosystems. This research provides actionable pathways through enhanced taxonomy and automated detection capabilities, but most critically demonstrates that securing infrastructure automation requires moving beyond traditional code security to understand how different IaC ecosystems inherently shape security risks. The systematic integration of security-aware development practices is not merely beneficial, it is essential for the secure evolution of modern infrastructure.

\section{Data Availability}
In the interest of transparency and reproducibility, we provide the complete set of artifacts necessary to replicate and validate our findings. These artifacts are hosted in an anonymized GitHub repository and include the source code with full implementation of our experiments, our datasets of IaC scripts, the experiment pipelines to reproduce the entire data collection, all results outputs of all experiments, and the setup instructions to enable straightforward reproduction.  
We have organized the resources into the following anonymous GitHub repository: \url{https://anonymous.4open.science/r/Taxonomy-of-smells-6E6F}.

\balance
\bibliographystyle{IEEEtran} 
\bibliography{references.bib}


\end{document}